\documentclass[aps,prl,twocolumn,showpacs,preprintnumbers,floatfix]{revtex4-1}

\usepackage{color}
\usepackage{graphicx}
\usepackage{bm}

\begin{document}

\title{Magnetically tunable Feshbach resonances in Li + Yb($^3P_J$)}

\author{Maykel L. Gonz\'alez-Mart\'inez}
\author{Jeremy M. Hutson}
\email{J.M.Hutson@durham.ac.uk} \affiliation{Joint Quantum Centre (JQC)
Durham/Newcastle, Department of Chemistry, Durham University, South Road,
Durham, DH1~3LE, United Kingdom}

\date{\today}

\begin{abstract}
We have investigated magnetically tunable Feshbach resonances arising from the
interaction of Li($^2S$) with metastable Yb($^3P_2$) and ($^3P_0$). For
Yb($^3P_2$), all the resonance features are strongly suppressed by inelastic
collisions that produce Yb in its lower-lying $^3P_1$ and $^3P_0$ states. For
Yb($^3P_0$), sharp resonances exist but they are extremely narrow (widths less
than 1~mG).
\end{abstract}

\maketitle

There is currently great interest in the production of ultracold molecules,
driven by potential applications in fields ranging from high-precision
measurement to ultracold chemistry \cite{Carr:NJPintro:2009}. Many groups have
succeeded in producing alkali-metal dimers in high-lying vibrational states by
either magnetoassociation or photoassociation \cite{Hutson:IRPC:2006,
Kohler:RMP:2006, Jones:RMP:2006}, and a few species have been transferred to
their absolute ground states, either incoherently by absorbtion followed by
spontaneous emission \cite{Sage:2005, Deiglmayr:2008} or coherently by
stimulated Raman adiabatic passage (STIRAP) \cite{Lang:ground:2008,
Ni:KRb:2008, Danzl:ground:2010, Aikawa:2010}. KRb molecules produced by STIRAP
\cite{Ni:KRb:2008} have been used to investigate ultracold chemical reactions
\cite{Ospelkaus:react:2010, Ni:2010} and the properties of dipolar quantum
gases \cite{Ni:2010}.

The alkali-metal dimers all have singlet ground states. There is great interest
in extending molecule formation to molecules with doublet ground states, such
as those formed from an alkali-metal or other closed-shell atom and an
alkaline-earth atom. Such molecules have properties that may be important in
quantum information processing \cite{Micheli:2006}. \.Zuchowski {\em et al.}\
\cite{Zuchowski:RbSr:2010} have shown that such systems can have magnetically
tunable Feshbach resonances, with the incoming channel coupled to a bound state
by the very weak distance-dependence of the hyperfine coupling. The resulting
Feshbach resonances are very narrow \cite{Brue:LiYb:2012, Brue:AlkYb:2013}, but
nevertheless several groups have begun experiments aimed at observing them in
systems such as Li+Yb \cite{Okano:2009, Hansen:2011, Ivanov:2011} and Rb+Yb
\cite{Baumer:2011, Muenchow:2011, Muenchow:thesis:2012}.

The Li+Yb system has particularly narrow resonances when the atoms are in their
ground states. Five of the seven stable isotopes of Yb have spin-zero nuclei,
and for these the resonances are predicted to be only a few $\mu$G wide
\cite{Brue:LiYb:2012}. The fermionic isotopes $^{171}$Yb and $^{173}$Yb are
predicted to have somewhat wider resonances, but even these are predicted to be
only around 1 mG wide \cite{Brue:LiYb:2012}. However, ultracold Yb can also be
prepared in its metastable $^3P_2$ state \cite{Yamaguchi:2008}, which has a
radiative lifetime of at least 15~s \cite{Mishra:2001, Yamaguchi:2008}. Atoms
in $P$ states are anisotropic \cite{Reid:1969}, so the interaction of
Yb($^3P_2$) with Li($^2S$) introduces several additional couplings that may be
expected to produce broader resonances \cite{Krems:atoms:2004}. Hansen {\em et
al.}\ \cite{Hansen:2013} have suggested using these for molecule formation. If
this can be achieved, it will open up a new route to the production of
molecules with both electron spin and electric dipole moments, which may be
applicable to a wide variety of species. The purpose of the present paper is to
investigate the feasibility of this approach.

\begin{figure}
\includegraphics[width=\linewidth]{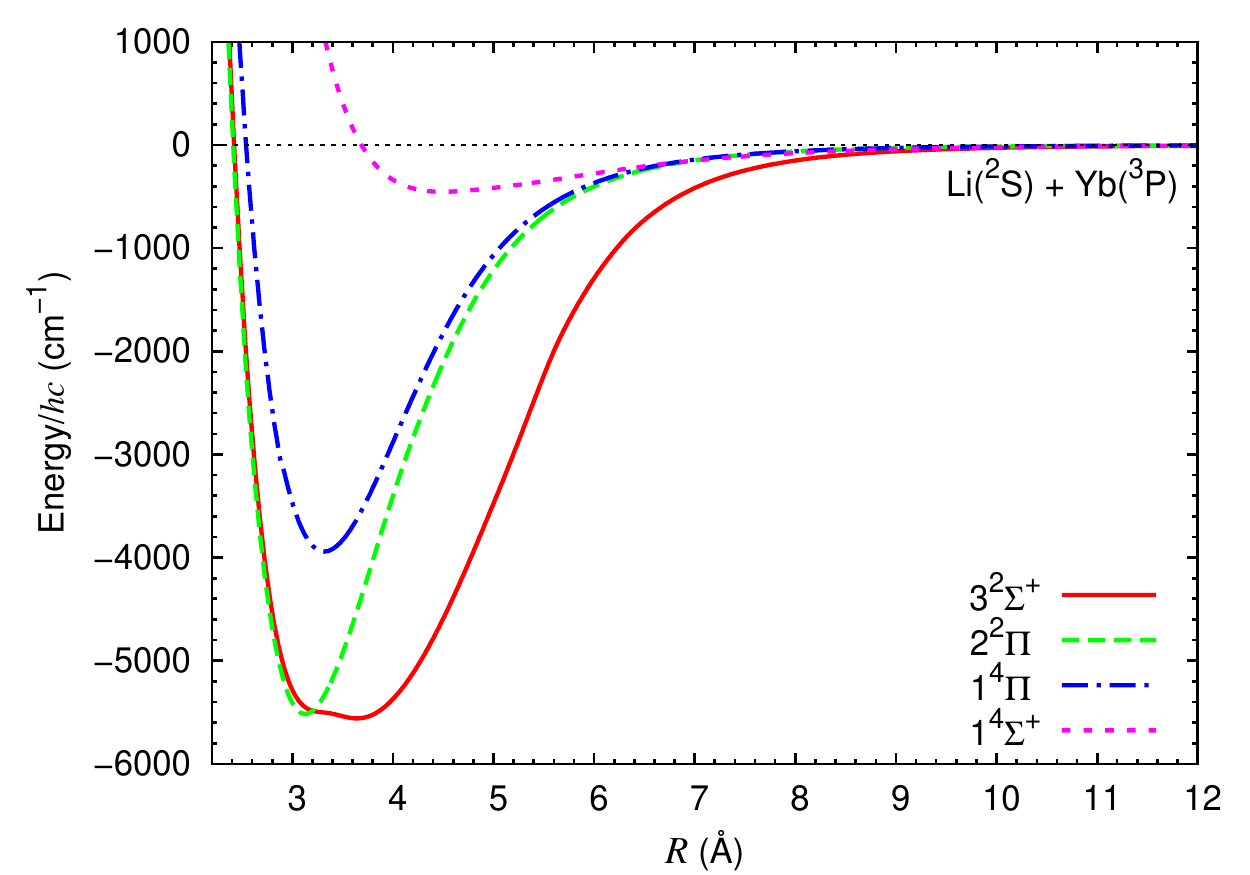}
\caption{Potential curves used in the present work for Li($^2S$) interacting
with Yb($^3P$), based on the electronic structure calculations of Gopakumar
{\em et al.}\ \cite{Gopakumar:2010}.} \label{fig:pot}
\end{figure}

Yb($^3P$) interacts with Li($^2S$) to produce four electronic states, of
$^2\Sigma^+$, $^2\Pi$, $^4\Sigma^+$ and $^4\Pi$ symmetry. In the present work
we have used potential curves for these states (neglecting spin-orbit coupling)
calculated by Gopakumar {\em et al.}\ \cite{Gopakumar:2010} using CASPT2
calculations (complete active space with second-order perturbation theory).
These are qualitatively similar to the curves of Zhang {\em et al.}\
\cite{Zhang:2010}. The two doublet states are each over 5000 cm$^{-1}$ deep,
while the quartet states are shallower. The $^2\Sigma^+$ state shows strong
attraction at considerably longer range than $^2\Pi$ because of chemical
bonding involving the Li($2s$) and Yb($6p_z$) orbitals. We have interpolated
the curves using the Reproducing Kernel Hilbert Space (RKHS) approach of Ho and
Rabitz \cite{Ho:1996}, and constrained them at long range to have $C_6$
coefficients that are the same for doublet and quartet curves but different for
$\Sigma$ and $\Pi$ curves. We obtained the value $C_6^0=2312.6\ E_{\rm h}a_0^6$
for Li($^2S$) + Yb($^3P$), using Tang's combination rule \cite{Tang:1969} with
the values of the static polarizability and dispersion coefficients for Li
\cite{Derevianko:2010} and Yb($^3P$) \cite{Dzuba:2010}. This was interpreted as
the average value, $C_6^0=(1/3)(C_6^\Sigma+2C_6^\Pi)$. The difference
$C_6^\Sigma-C_6^\Pi$ is not known for Li+Yb, so we used the approximation that
the ratio $C_6^\Sigma/C_6^\Pi$ is the same for LiYb as for LiSr, where Jiang
{\em et al.}\ \cite{Jiang:2013} obtained the ratio 1.146; this gives
$C_6^\Sigma=2528$ and $C_6^\Pi=2205\ E_{\rm h}a_0^6$ for Li($^2S$) + Yb($^3P$).
The resulting curves are shown in Fig.~\ref{fig:pot} \footnote{To obtain smooth
curves with the correct long-range behavior, we omitted {\em ab initio} points
outside $R=13.0, 8.6, 9.6$ and $8.6\,a_0$ for the $^2\Sigma^+$, $^2\Pi$,
$^4\Sigma^+$ and $^4\Pi$ curves, respectively, and between 9.6 and $10.5\,a_0$
for the $^2\Sigma^+$ curve.}.

We have carried out coupled-channel scattering calculations to characterize the
magnetically tunable Feshbach resonances. The collision Hamiltonian is
\begin{equation}
\frac{\hbar^2}{2\mu}\left(-\frac{d^2}{dR^2}+\frac{\hat L^2}{R^2}\right) + \hat H_{\rm Li}
+ \hat H_{\rm Yb} + \hat U(R), \label{eq:ham}
\end{equation}
where $R$ is the internuclear distance, $\mu$ is the reduced mass, and $\hat L$
is the angular momentum operator for relative motion of the two atoms. The
free-atom Hamiltonians are taken to be
\begin{eqnarray}
\hat H_{\rm Li} &=& \zeta_{\rm Li}\hat \imath_{\rm Li}\cdot\hat s_{\rm Li} +
\left(g_{e}\mu_{\rm B}\hat s_{z,{\rm Li}}
+ g_{\rm Li}\mu_{\rm N}\hat \imath_{z,{\rm Li}}\right)B;\cr
\hat H_{\rm Yb} &=& a^{\rm so}_{\rm Yb}\hat l \cdot\hat s_{\rm Yb}
+ \left(\mu_{\rm B}\hat l_z + g_{e}\mu_{\rm B}\hat s_{z,{\rm Yb}}\right)B.
\end{eqnarray}
Here we use the convention that quantum numbers of the individual collision
partners are represented by lower-case letters, so that $\hat l$ and $\hat
s_{\rm Yb}$ are the orbital and electron spin angular momentum operators for
the Yb atom, and $\hat s_{\rm Li}$ and $\hat \imath_{\rm Li}$ are the electron
and nuclear spin operators for the Li atom. $\zeta_{\rm Li}$ is the hyperfine
coupling constant for Li($^2S$), $a^{\rm so}_{\rm Yb}$ is the spin-orbit
coupling constant for Yb($^3P$), and $B$ is the magnetic field. In the present
work we used $a^{\rm so}_{\rm Yb}=859.1905\times hc$ cm$^{-1}$, which gives the
correct value for the splitting between the $^3P_2$ and $^3P_1$ states
\cite{Meggers:1978}.  The interaction operator $\hat U(R)$ may be written
\begin{equation}
\hat U(R) = \sum_{\Lambda,S} | \Lambda,S \rangle V^{\Lambda,S}(R) \langle \Lambda,S\ |
+ \hat V^{\rm d}(R),
\label{eq:inter}
\end{equation}
where $\Lambda=0$ and 1 indicates the $\Sigma$ and $\Pi$ states, $S=1/2$ and
$3/2$ is the total electron spin (for the doublet and quartet potentials,
respectively), and $\hat V^{\rm d}(R)$ represents the dipolar interaction
between the magnetic moments due to Li and Yb unpaired electrons.

We have implemented this Hamiltonian in the BOUND program for calculating
bound-state energies \cite{Hutson:bound:2011} and the MOLSCAT scattering
package \cite{molscat:v14-short}, using two different basis sets: $|ls_{\rm
Yb}jm_j\rangle |s_{\rm Li}m_{s,{\rm Li}} \rangle |i_{\rm Li}m_{i,{\rm Li}}
\rangle |LM_L\rangle$ and $|lm_ls_{\rm Yb} m_{s,{\rm Yb}}\rangle |s_{\rm
Li}m_{s,{\rm Li}}\rangle |i_{\rm Li}m_{i,{\rm Li}}\rangle |LM_L\rangle$. Here
$j=0$, 1 or 2 is the total angular momentum of Yb and the $m$ and $M$ quantum
numbers are angular momentum projections onto the axis of the applied magnetic
field. The only rigorously conserved quantities are the projection of the total
angular momentum, $M_{\rm tot}=m_j+m_{s,{\rm Li}}+m_{i,{\rm Li}}+M_L =
m_l+m_{s,{\rm Yb}}+m_{s,{\rm Li}}+m_{i,{\rm Li}}+M_L$ and total parity
$P=(-1)^{L+1}$. We have verified that the two basis sets give identical results
when all possible values of $j$ and the projection quantum numbers for a given
$M_{\rm tot}$ are included. However, the first of the two basis sets has the
advantage that it is possible to restrict the basis functions to those
correlating with an individual spin-orbit state of Yb, which will be important
in the discussion below.

\begin{figure}
\includegraphics[width=\linewidth]{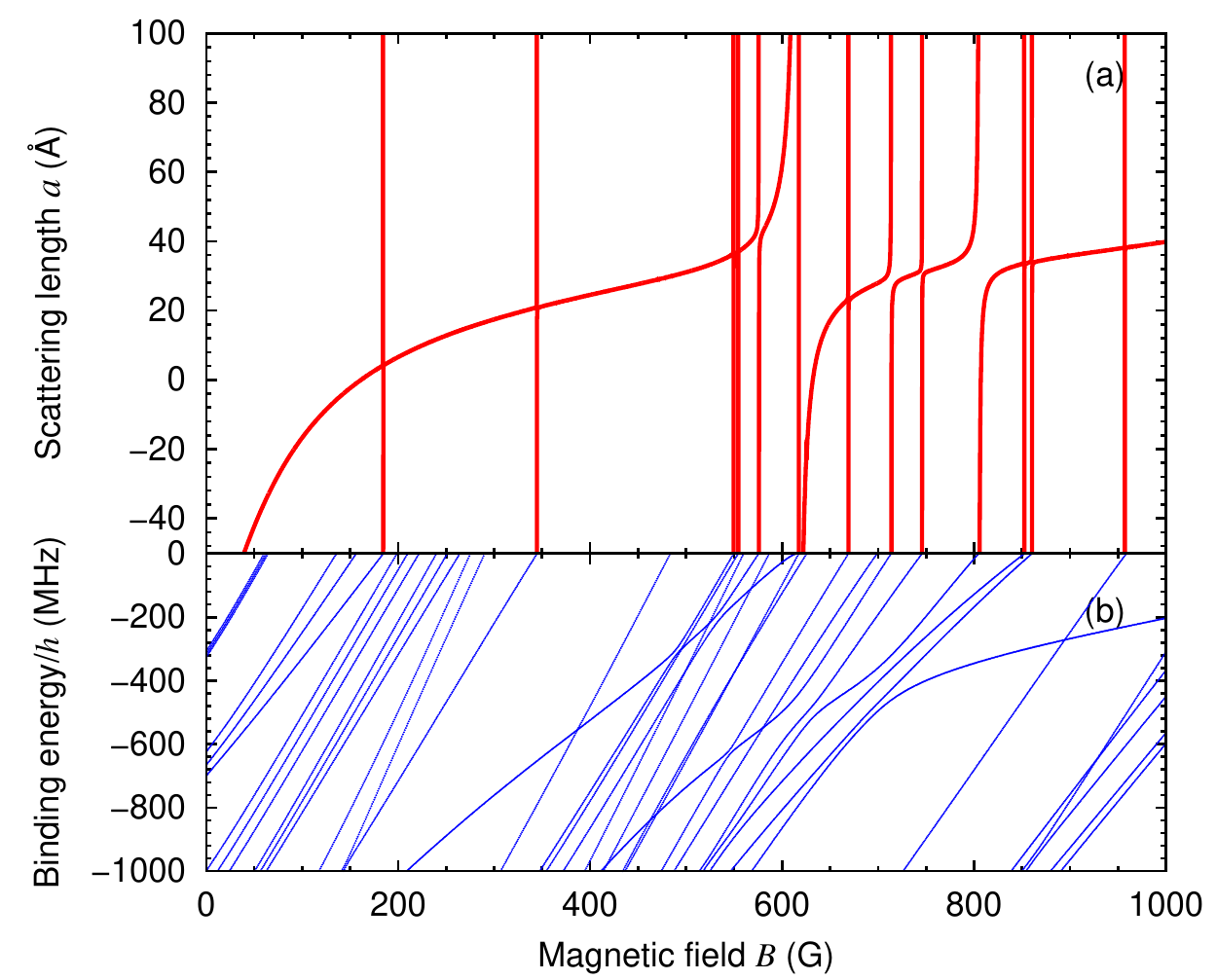}
\caption{(a) Scattering length $a(B)$ for collisions of Li in its absolute
ground state with Yb($^3P_2,m_j=-2$), including only Yb basis functions with
$j=2$. (b) The near-threshold bound states responsible for the resonances in
(a).} \label{fig:j2only}
\end{figure}

We first consider calculations for Li($^2S$) + Yb($^3P_2$) with both $^6$Li and
$^{174}$Yb in their lowest Zeeman state ($f=1/2,m_f=1/2$ for Li and $m_j=-2$
for Yb), restricting the basis set to functions with $j=2$. Convergence was
achieved with $L_{\rm max}=12$. Fig.~\ref{fig:j2only}(a) shows the resulting
s-wave scattering length $a(B)$, calculated at a collision energy of 100~nK
$\times\ k_{\rm B}$, as a function of magnetic field $B$.
It may be seen that there are numerous resonances with widths
in the range 0.1 to 10 G, which at first sight look promising candidates for
molecule formation. Fig.~\ref{fig:j2only}(b) shows the near-threshold bound
states calculated with the same basis set. It may be seen that only a small
fraction of the bound states that cross threshold cause visible resonances; the
widest resonances are due to bound states that are dominated by low $L$ quantum
numbers.

\begin{figure}
\includegraphics[width=\linewidth]{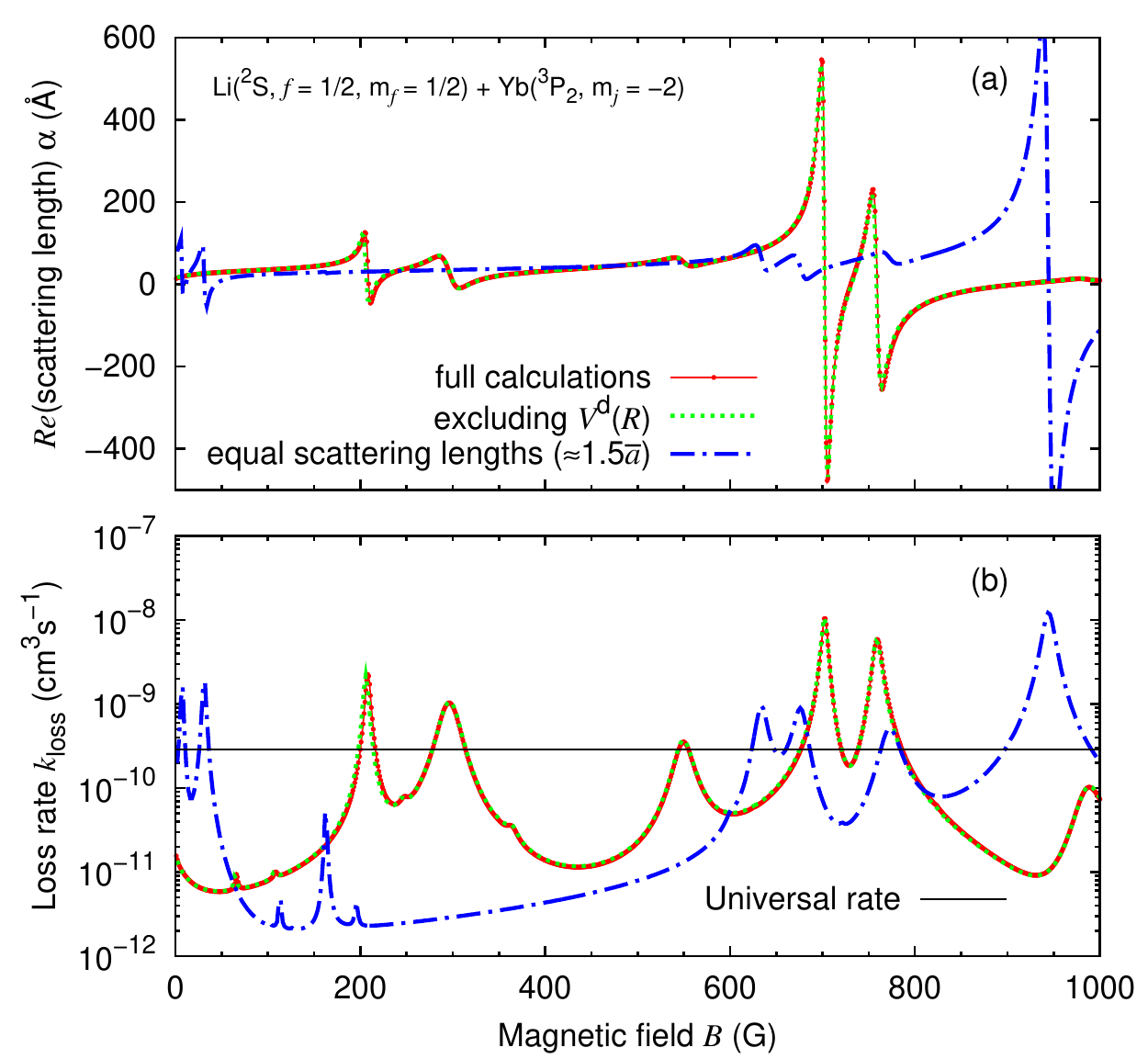}
\caption{(a) The real part $\alpha(B)$ of the scattering length for collisions of Li
in its absolute ground state with Yb($^3P_2,m_j=-2$), including Yb basis
functions with $j=0$, 1 and 2. (b) The corresponding 2-body loss rate $k_{\rm
loss}$.} \label{fig:allj-3/2}
\end{figure}

The promising results shown in Fig.~\ref{fig:j2only} unfortunately neglect
couplings to the lower-lying  $^3P_0$ and $^3P_1$ states of Yb. It is known
that Feshbach resonances in the presence of inelastic scattering have
signatures that are no longer pole-like, but instead exhibit more complicated
lineshapes in which the poles are suppressed \cite{Hutson:res:2007}. The
scattering length in the presence of inelastic scattering is complex,
$a(B)=\alpha(B)-{\rm i}\beta(B)$, where $\beta(B)$ represents a 2-body
inelastic loss rate $k_{\rm loss}\approx 4\pi\hbar\beta(B)/\mu$. We have
therefore repeated the scattering calculations for Li+Yb($^3P_2,m_j=-2)$,
including the $^3P_0$ and $^3P_1$ basis functions, and produced the scattering
length and loss rate shown as red curves in Fig.~\ref{fig:allj-3/2}. It may be
seen that the inelastic processes have greatly reduced the amplitude of the
oscillations in $a(B)$ and produced fast loss rates over most of the range of
magnetic field considered. Such fast loss rates are likely to prevent the use
of these resonances for molecule formation.

The loss rates shown in Fig.\ \ref{fig:allj-3/2} are generally lower than the
``universal" loss rate predicted by Idziaszek and Julienne
\cite{Idziaszek:PRL:2010} for systems in which all collisions that reach short
range produce inelasticity, which is $2.9\times10^{-10}$ cm$^3$ s$^{-1}$ in the
present case, shown by the horizontal black line in Fig.\ \ref{fig:allj-3/2}.
However, they are considerably faster than those observed experimentally for
Yb($^3P_2,m_j=-2$)+Yb($^1S_0$) \cite{Uetake:2012}, which are below $10^{-12}$
cm$^3$ s$^{-1}$ at fields up to 1~G. Interestingly, repeating the
Li+Yb($^3P_2,m_j=-2)$ calculations with the Li spins set to zero also produces
much lower loss rates, below $10^{-12}$ cm$^3$ s$^{-1}$ except near narrow
resonances using the doublet potential curves, and even lower using the less
anisotropic quartet potentials. This makes it clear that the {\em difference}
between the doublet and quartet potentials, which can drive spin exchange
processes, is key in causing the fast loss rates.

It should be emphasized that the inelastic transitions from the $j=2$, $m_j=-2$
state are driven principally by the Born-Oppenheimer potentials of
Fig.~\ref{fig:pot} and {\em not} to any great extent by the magnetic dipolar
interaction $V^{\rm d}(R)$. To demonstrate this we have repeated the
calculation with the magnetic dipolar term omitted, and obtained the results
shown as dotted green curves in Fig.~\ref{fig:allj-3/2}. These differ only
slightly from the results with $V^{\rm d}(R)$ included. When states of $\Sigma$
and $\Pi$ character both exist and have different potential curves, the
difference may be viewed as an ``anisotropy" in the interaction, and produces
matrix elements off-diagonal in both $j$ and $m_j$. In the present case, the
state-to-state cross sections for formation of Yb($^3P_1$) are approximately a
factor of 100 larger than those for formation of Yb($^3P_0$).

\begin{figure}
\includegraphics[width=\linewidth]{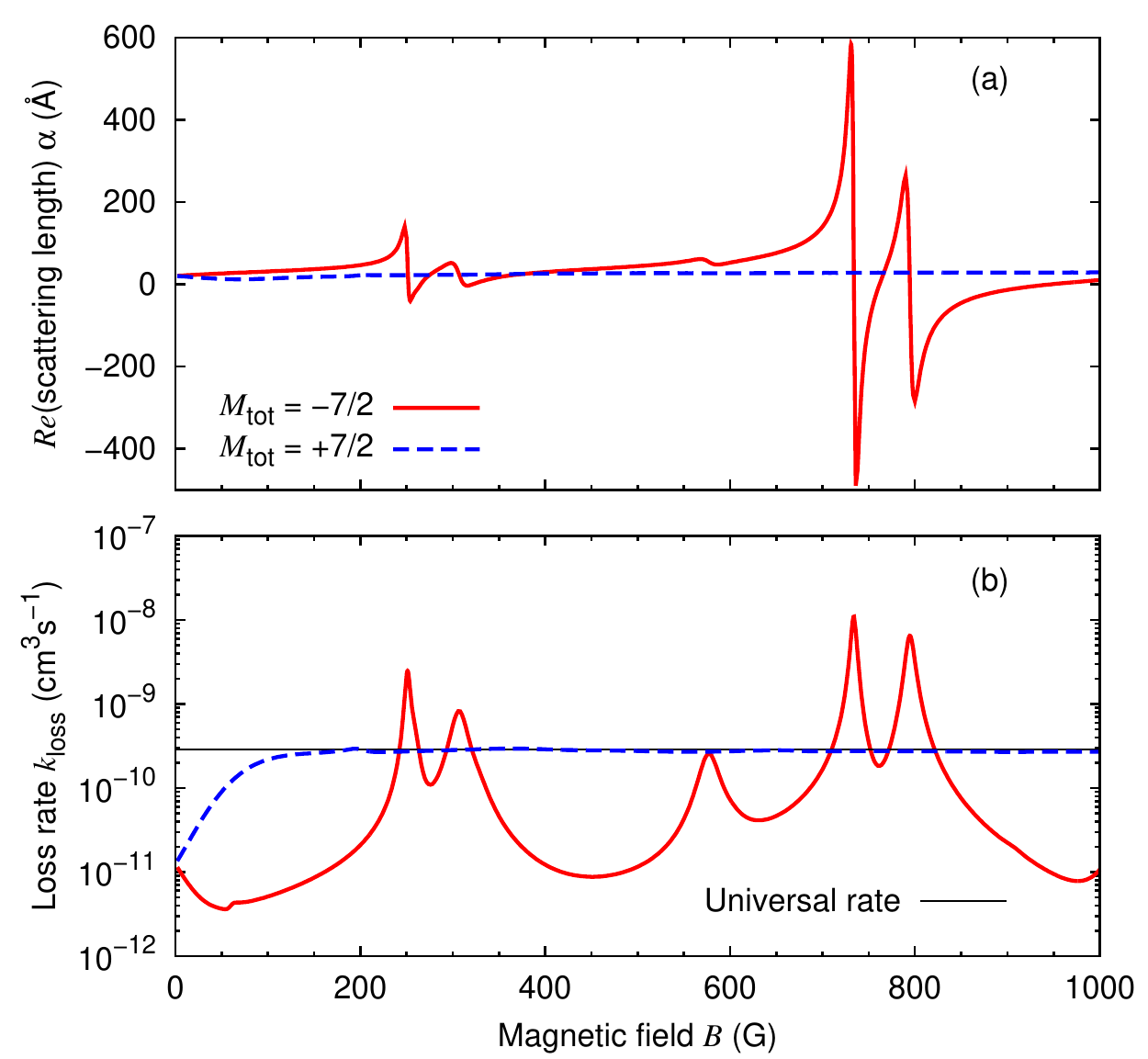}
\caption{(a) The real part $\alpha(B)$ of the scattering length for spin-stretched
collisions of Li and Yb($^3P_2$), $M_{\rm tot}=-7/2$ and $M_{\rm tot}=+7/2$,
including Yb basis functions with $j=0$, 1 and 2. (b) The corresponding two-body
loss rates $k_{\rm loss}$.} \label{fig:allj-7/2}
\end{figure}

We have also investigated resonances for spin-stretched collisions,
Yb($^3P_2,m_j=-2$)+Li($m_f=-3/2$) ($M_{\rm tot}=-7/2$) and
Yb($^3P_2,m_j=+2$)+Li($m_f=+3/2$) ($M_{\rm tot}=+7/2$). The results are shown
in Fig.~\ref{fig:allj-7/2}. The results for $M_{\rm tot}=-7/2$ are remarkably
similar to those for $M_{\rm tot}=-3/2$ in Fig.~\ref{fig:j2only}, with just a
small shift in field: the $-3/2$ and $-7/2$ thresholds are close in energy
(within 200~MHz), the resonances are caused by the same bound states, and the
principal loss mechanism (formation of Yb($^3P_1$)) is the same. For $M_{\rm
tot}=+7/2$, by contrast, spin relaxation to form lower-lying Zeeman states of
Yb($^3P_2$) is the dominant mechanism. These spin relaxation processes are so
fast that the loss rate is very close to the ``universal" rate at fields above
100~G. The situation is thus quite different from that for the alkali-metal
pairs, where inelastic processes are strongly suppressed for spin-stretched
collisions. The difference arises because, for the alkali-metal pairs, the
interaction potential is isotropic and only the magnetic dipole interaction can
cause spin relaxation. In the present case, by contrast, changes in $j$ and
$m_j$ can be driven directly by the anisotropic interaction potentials.

The potential energy curves of Gopakumar {\em et al.}\ \cite{Gopakumar:2010}
are likely to be qualitatively correct, but they are unlikely to be accurate
enough to predict the correct values of the four scattering lengths. However,
the overall density of levels is controlled by the $C_6$ coefficient, and (no
matter what the scattering length), every threshold supports an s-wave bound
state within approximately 7~GHz of threshold. There are also higher-$L$ states
supported by these and deeper levels. As may be seen in Fig.\ \ref{fig:j2only},
the bound states are actually well distributed across this range, and many of
them cross the $m_j=-2$ threshold at fields below 1000~G. The potentials we
have used should therefore reliably predict the overall {\em density} of
Feshbach resonances, but will not give quantitative predictions of their
positions and widths.

It is important to consider whether the potential curves will accurately
predict the {\em strength of decay} and hence the extent of the suppression of
the peaks in the scattering length. In particular, for the alkali-metal dimers
it is known that the rate of inelastic (spin exchange) collisions depends
strongly on the {\em difference} in scattering lengths between the singlet and
triplet states \cite{Julienne:1997}, and that inelastic rates can be
anomalously low when the singlet and triplet scattering lengths are very
similar. The particular potential curves used in Fig.~\ref{fig:j2only} have
scattering lengths of $-$89.27 and 50.30~$a_0$ for the $^2\Sigma^+$ and
$^4\Sigma^+$ states and 58.30 and 90.82~$a_0$ for the $^2\Pi$ and $^4\Pi$
states. In order to estimate whether an analogous effect may reduce the loss
rates in Li + Yb($^3P_2$), we have investigated an artificial problem in which
the potentials are adjusted so that all four scattering lengths have the same
value, 57.75~$a_0$, chosen to be approximately 1.5 times the mean scattering
length $\bar{a}$ \cite{Gribakin:1993}, which is about 40~$a_0$ for this system.
The results are shown as blue dot-dashed curves in Fig.~\ref{fig:allj-3/2}; the
resonances are of course in different locations, but the degree of damping is
comparable. This demonstrates that equal scattering lengths are not sufficient
to suppress inelasticity in the present case. The suppression observed for spin
exchange in $^{87}$Rb in ref.\ \cite{Julienne:1997} was for inelastic
collisions involving two near-threshold channels, where similar scattering
lengths implied similar long-range wavefunctions. In the present case, with a
large kinetic energy release, the long-range wavefunctions are quite different
even when the scattering lengths are the same.

\begin{figure}
\includegraphics[width=\linewidth]{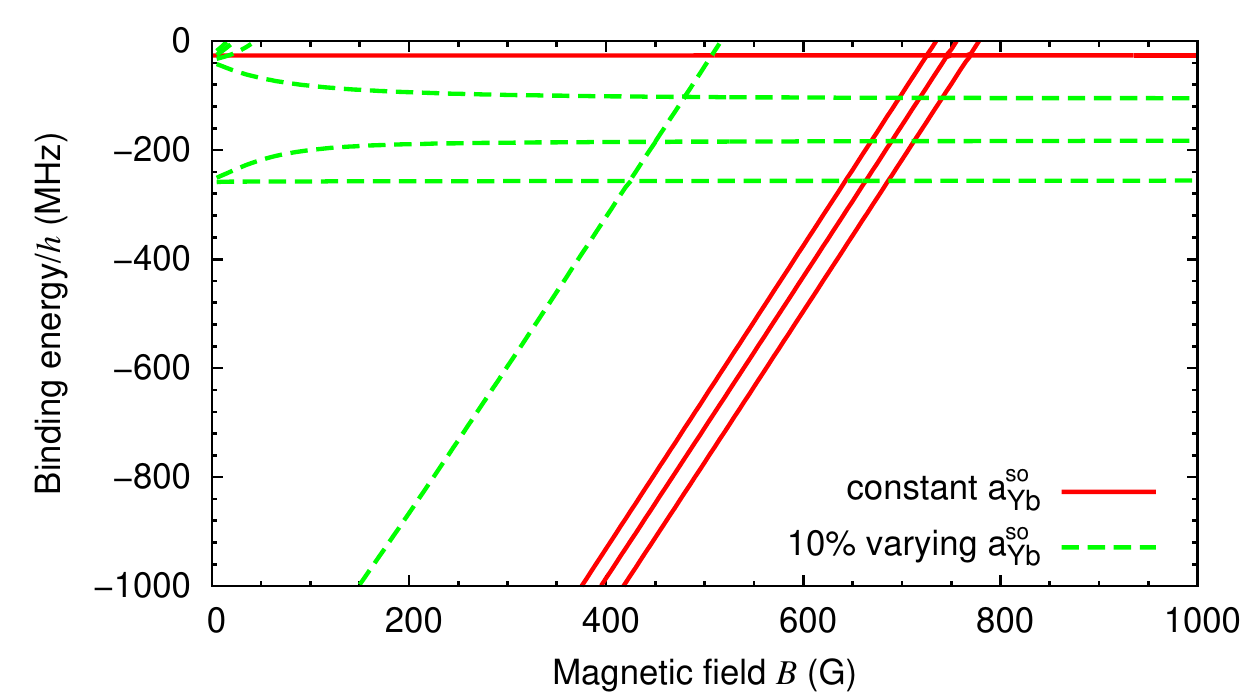}
\caption{Near-threshold bound states relative to the absolute ground state Li +
Yb($^3P_0)$ threshold, with the spin-orbit coupling constant held fixed at its atomic
value (solid, red) and allowed to vary by 10\% across the range of chemical interactions
(dashed, green).}
\label{fig:j0}
\end{figure}

We have also explored whether the $^3P_0$ metastable state of Yb offers the
possibility of broad Feshbach resonances. This state has a radiative lifetime
that is of order 20~s for $^{171}$Yb and $^{173}$Yb \cite{Porsev:2004,
Hoyt:2005, Hong:2005} and is too long to be observed without an applied field
for bosonic Yb isotopes \cite{Barber:2006}. The near-threshold bound states
that might cause resonances at this threshold (with Li in its absolute ground
state) are shown as red lines in Fig.~\ref{fig:j0}. There are bound states that
cross threshold as a function of magnetic field, arising from states in which
the Li is in a higher magnetic or hyperfine state. For the potentials
considered here, the states that cross threshold have $L=4$. They are coupled
to the incoming s-wave channel only by the spin dipolar term, acting in second
order, or indirectly by the potential anisotropy via the far-away $^3P_1$ and
$^3P_2$ states. The resulting Feshbach resonances are so narrow that we were
unable to locate them in scattering calculations.

The interaction operator (\ref{eq:inter}) does neglect some terms. In
particular, it neglects the $R$-dependence of the spin-orbit coupling operator.
This might in principle introduce stronger coupling between the bound and
continuum states, in the same way as the $R$-dependence of the hyperfine
coupling in systems such as RbSr \cite{Zuchowski:RbSr:2010} and alkali metal +
Yb($^1S$) \cite{Brue:LiYb:2012, Brue:AlkYb:2013}. We have therefore repeated
the calculations introducing an $R$-dependent spin-orbit coupling $a_{\rm
Yb}^{\rm so}(R)$ that varies by up to 20\% of its value across the well region,
with a shape obtained by scaling the electronic structure calculations of
Gopakumar {\em et al.}\ \cite{Gopakumar:2010}. The bound states for a 10\%
change are shown as dashed green lines in Fig.~\ref{fig:j0}; they naturally
shift in position, but the couplings between the bound states and the incoming
wave are not significantly larger and we were again unable to identify Feshbach
resonances in scattering calculations. The mechanism that produces Feshbach
resonances for Li+Yb($^1S$), through the $R$-dependence of the Li hyperfine
coupling \cite{Brue:LiYb:2012}, will still exist for Yb($^3P_0$), but it
nevertheless appears that resonances at the $^3P_0$ threshold are not
significantly wider than those for the ground state.

We thus conclude that Feshbach resonances arising from interaction of any atom
in a $^3P_2$ state with an alkali-metal atom are likely to be strongly
decaying, provided that the $^3P_0$ and $^3P_1$ states lie below the $^3P_2$
state. Resonances that occur at the $^3P_0$ threshold are extremely narrow.

\noindent [Note: We are aware that the group at Temple University
\cite{Kotochigova:private:2013} is carrying out a parallel investigation into
Li+Yb resonances at the Yb($^3P_2$) threshold.]

The authors are grateful to EPSRC and EOARD for funding.

\bibliography{../all}

\begin{thebibliography}{48}%
\makeatletter
\providecommand \@ifxundefined [1]{%
 \@ifx{#1\undefined}
}%
\providecommand \@ifnum [1]{%
 \ifnum #1\expandafter \@firstoftwo
 \else \expandafter \@secondoftwo
 \fi
}%
\providecommand \@ifx [1]{%
 \ifx #1\expandafter \@firstoftwo
 \else \expandafter \@secondoftwo
 \fi
}%
\providecommand \natexlab [1]{#1}%
\providecommand \enquote  [1]{``#1''}%
\providecommand \bibnamefont  [1]{#1}%
\providecommand \bibfnamefont [1]{#1}%
\providecommand \citenamefont [1]{#1}%
\providecommand \href@noop [0]{\@secondoftwo}%
\providecommand \href [0]{\begingroup \@sanitize@url \@href}%
\providecommand \@href[1]{\@@startlink{#1}\@@href}%
\providecommand \@@href[1]{\endgroup#1\@@endlink}%
\providecommand \@sanitize@url [0]{\catcode `\\12\catcode `\$12\catcode
  `\&12\catcode `\#12\catcode `\^12\catcode `\_12\catcode `\%12\relax}%
\providecommand \@@startlink[1]{}%
\providecommand \@@endlink[0]{}%
\providecommand \url  [0]{\begingroup\@sanitize@url \@url }%
\providecommand \@url [1]{\endgroup\@href {#1}{\urlprefix }}%
\providecommand \urlprefix  [0]{URL }%
\providecommand \Eprint [0]{\href }%
\providecommand \doibase [0]{http://dx.doi.org/}%
\providecommand \selectlanguage [0]{\@gobble}%
\providecommand \bibinfo  [0]{\@secondoftwo}%
\providecommand \bibfield  [0]{\@secondoftwo}%
\providecommand \translation [1]{[#1]}%
\providecommand \BibitemOpen [0]{}%
\providecommand \bibitemStop [0]{}%
\providecommand \bibitemNoStop [0]{.\EOS\space}%
\providecommand \EOS [0]{\spacefactor3000\relax}%
\providecommand \BibitemShut  [1]{\csname bibitem#1\endcsname}%
\let\auto@bib@innerbib\@empty
\bibitem [{\citenamefont {Carr}\ \emph {et~al.}(2009)\citenamefont {Carr},
  \citenamefont {{DeMille}}, \citenamefont {Krems},\ and\ \citenamefont
  {Ye}}]{Carr:NJPintro:2009}%
  \BibitemOpen
  \bibfield  {author} {\bibinfo {author} {\bibfnamefont {L.~D.}\ \bibnamefont
  {Carr}}, \bibinfo {author} {\bibfnamefont {D.}~\bibnamefont {{DeMille}}},
  \bibinfo {author} {\bibfnamefont {R.~V.}\ \bibnamefont {Krems}}, \ and\
  \bibinfo {author} {\bibfnamefont {J.}~\bibnamefont {Ye}},\ }\href@noop {}
  {\bibfield  {journal} {\bibinfo  {journal} {New J. Phys.}\ }\textbf {\bibinfo
  {volume} {11}},\ \bibinfo {pages} {055049} (\bibinfo {year}
  {2009})}\BibitemShut {NoStop}%
\bibitem [{\citenamefont {Hutson}\ and\ \citenamefont
  {Sold\'{a}n}(2006)}]{Hutson:IRPC:2006}%
  \BibitemOpen
  \bibfield  {author} {\bibinfo {author} {\bibfnamefont {J.~M.}\ \bibnamefont
  {Hutson}}\ and\ \bibinfo {author} {\bibfnamefont {P.}~\bibnamefont
  {Sold\'{a}n}},\ }\href@noop {} {\bibfield  {journal} {\bibinfo  {journal}
  {Int. Rev. Phys. Chem.}\ }\textbf {\bibinfo {volume} {25}},\ \bibinfo {pages}
  {497} (\bibinfo {year} {2006})}\BibitemShut {NoStop}%
\bibitem [{\citenamefont {K\"{o}hler}\ \emph {et~al.}(2006)\citenamefont
  {K\"{o}hler}, \citenamefont {Goral},\ and\ \citenamefont
  {Julienne}}]{Kohler:RMP:2006}%
  \BibitemOpen
  \bibfield  {author} {\bibinfo {author} {\bibfnamefont {T.}~\bibnamefont
  {K\"{o}hler}}, \bibinfo {author} {\bibfnamefont {K.}~\bibnamefont {Goral}}, \
  and\ \bibinfo {author} {\bibfnamefont {P.~S.}\ \bibnamefont {Julienne}},\
  }\href@noop {} {\bibfield  {journal} {\bibinfo  {journal} {Rev. Mod. Phys.}\
  }\textbf {\bibinfo {volume} {78}},\ \bibinfo {pages} {1311} (\bibinfo {year}
  {2006})}\BibitemShut {NoStop}%
\bibitem [{\citenamefont {Jones}\ \emph {et~al.}(2006)\citenamefont {Jones},
  \citenamefont {Tiesinga}, \citenamefont {Lett},\ and\ \citenamefont
  {Julienne}}]{Jones:RMP:2006}%
  \BibitemOpen
  \bibfield  {author} {\bibinfo {author} {\bibfnamefont {K.~M.}\ \bibnamefont
  {Jones}}, \bibinfo {author} {\bibfnamefont {E.}~\bibnamefont {Tiesinga}},
  \bibinfo {author} {\bibfnamefont {P.~D.}\ \bibnamefont {Lett}}, \ and\
  \bibinfo {author} {\bibfnamefont {P.~S.}\ \bibnamefont {Julienne}},\
  }\href@noop {} {\bibfield  {journal} {\bibinfo  {journal} {Rev. Mod. Phys.}\
  }\textbf {\bibinfo {volume} {78}},\ \bibinfo {pages} {483} (\bibinfo {year}
  {2006})}\BibitemShut {NoStop}%
\bibitem [{\citenamefont {Sage}\ \emph {et~al.}(2005)\citenamefont {Sage},
  \citenamefont {Sainis}, \citenamefont {Bergeman},\ and\ \citenamefont
  {DeMille}}]{Sage:2005}%
  \BibitemOpen
  \bibfield  {author} {\bibinfo {author} {\bibfnamefont {J.~M.}\ \bibnamefont
  {Sage}}, \bibinfo {author} {\bibfnamefont {S.}~\bibnamefont {Sainis}},
  \bibinfo {author} {\bibfnamefont {T.}~\bibnamefont {Bergeman}}, \ and\
  \bibinfo {author} {\bibfnamefont {D.}~\bibnamefont {DeMille}},\ }\href@noop
  {} {\bibfield  {journal} {\bibinfo  {journal} {Phys. Rev. Lett.}\ }\textbf
  {\bibinfo {volume} {94}},\ \bibinfo {pages} {203001} (\bibinfo {year}
  {2005})}\BibitemShut {NoStop}%
\bibitem [{\citenamefont {Deiglmayr}\ \emph {et~al.}(2008)\citenamefont
  {Deiglmayr}, \citenamefont {Grochola}, \citenamefont {Repp}, \citenamefont
  {M\"ortlbauer}, \citenamefont {Gl\"uck}, \citenamefont {Lange}, \citenamefont
  {Dulieu}, \citenamefont {Wester},\ and\ \citenamefont
  {Weidem\"uller}}]{Deiglmayr:2008}%
  \BibitemOpen
  \bibfield  {author} {\bibinfo {author} {\bibfnamefont {J.}~\bibnamefont
  {Deiglmayr}}, \bibinfo {author} {\bibfnamefont {A.}~\bibnamefont {Grochola}},
  \bibinfo {author} {\bibfnamefont {M.}~\bibnamefont {Repp}}, \bibinfo {author}
  {\bibfnamefont {K.}~\bibnamefont {M\"ortlbauer}}, \bibinfo {author}
  {\bibfnamefont {C.}~\bibnamefont {Gl\"uck}}, \bibinfo {author} {\bibfnamefont
  {J.}~\bibnamefont {Lange}}, \bibinfo {author} {\bibfnamefont
  {O.}~\bibnamefont {Dulieu}}, \bibinfo {author} {\bibfnamefont
  {R.}~\bibnamefont {Wester}}, \ and\ \bibinfo {author} {\bibfnamefont
  {M.}~\bibnamefont {Weidem\"uller}},\ }\href@noop {} {\bibfield  {journal}
  {\bibinfo  {journal} {Phys. Rev. Lett.}\ }\textbf {\bibinfo {volume} {101}},\
  \bibinfo {pages} {133004} (\bibinfo {year} {2008})}\BibitemShut {NoStop}%
\bibitem [{\citenamefont {Lang}\ \emph {et~al.}(2008)\citenamefont {Lang},
  \citenamefont {Winkler}, \citenamefont {Strauss}, \citenamefont {Grimm},\
  and\ \citenamefont {Hecker~Denschlag}}]{Lang:ground:2008}%
  \BibitemOpen
  \bibfield  {author} {\bibinfo {author} {\bibfnamefont {F.}~\bibnamefont
  {Lang}}, \bibinfo {author} {\bibfnamefont {K.}~\bibnamefont {Winkler}},
  \bibinfo {author} {\bibfnamefont {C.}~\bibnamefont {Strauss}}, \bibinfo
  {author} {\bibfnamefont {R.}~\bibnamefont {Grimm}}, \ and\ \bibinfo {author}
  {\bibfnamefont {J.}~\bibnamefont {Hecker~Denschlag}},\ }\href@noop {}
  {\bibfield  {journal} {\bibinfo  {journal} {Phys. Rev. Lett.}\ }\textbf
  {\bibinfo {volume} {101}},\ \bibinfo {pages} {133005} (\bibinfo {year}
  {2008})}\BibitemShut {NoStop}%
\bibitem [{\citenamefont {Ni}\ \emph {et~al.}(2008)\citenamefont {Ni},
  \citenamefont {Ospelkaus}, \citenamefont {{de Miranda}}, \citenamefont
  {Pe'er}, \citenamefont {Neyenhuis}, \citenamefont {Zirbel}, \citenamefont
  {Kotochigova}, \citenamefont {Julienne}, \citenamefont {Jin},\ and\
  \citenamefont {Ye}}]{Ni:KRb:2008}%
  \BibitemOpen
  \bibfield  {author} {\bibinfo {author} {\bibfnamefont {K.-K.}\ \bibnamefont
  {Ni}}, \bibinfo {author} {\bibfnamefont {S.}~\bibnamefont {Ospelkaus}},
  \bibinfo {author} {\bibfnamefont {M.~H.~G.}\ \bibnamefont {{de Miranda}}},
  \bibinfo {author} {\bibfnamefont {A.}~\bibnamefont {Pe'er}}, \bibinfo
  {author} {\bibfnamefont {B.}~\bibnamefont {Neyenhuis}}, \bibinfo {author}
  {\bibfnamefont {J.~J.}\ \bibnamefont {Zirbel}}, \bibinfo {author}
  {\bibfnamefont {S.}~\bibnamefont {Kotochigova}}, \bibinfo {author}
  {\bibfnamefont {P.~S.}\ \bibnamefont {Julienne}}, \bibinfo {author}
  {\bibfnamefont {D.~S.}\ \bibnamefont {Jin}}, \ and\ \bibinfo {author}
  {\bibfnamefont {J.}~\bibnamefont {Ye}},\ }\href@noop {} {\bibfield  {journal}
  {\bibinfo  {journal} {Science}\ }\textbf {\bibinfo {volume} {322}},\ \bibinfo
  {pages} {231} (\bibinfo {year} {2008})}\BibitemShut {NoStop}%
\bibitem [{\citenamefont {Danzl}\ \emph {et~al.}(2010)\citenamefont {Danzl},
  \citenamefont {Mark}, \citenamefont {Haller}, \citenamefont {Gustavsson},
  \citenamefont {Hart}, \citenamefont {Aldegunde}, \citenamefont {Hutson},\
  and\ \citenamefont {N\"agerl}}]{Danzl:ground:2010}%
  \BibitemOpen
  \bibfield  {author} {\bibinfo {author} {\bibfnamefont {J.~G.}\ \bibnamefont
  {Danzl}}, \bibinfo {author} {\bibfnamefont {M.~J.}\ \bibnamefont {Mark}},
  \bibinfo {author} {\bibfnamefont {E.}~\bibnamefont {Haller}}, \bibinfo
  {author} {\bibfnamefont {M.}~\bibnamefont {Gustavsson}}, \bibinfo {author}
  {\bibfnamefont {R.}~\bibnamefont {Hart}}, \bibinfo {author} {\bibfnamefont
  {J.}~\bibnamefont {Aldegunde}}, \bibinfo {author} {\bibfnamefont {J.~M.}\
  \bibnamefont {Hutson}}, \ and\ \bibinfo {author} {\bibfnamefont {H.-C.}\
  \bibnamefont {N\"agerl}},\ }\href {\doibase doi:10.1038/nphys1533} {\bibfield
   {journal} {\bibinfo  {journal} {Nature Phys.}\ }\textbf {\bibinfo {volume}
  {6}},\ \bibinfo {pages} {265} (\bibinfo {year} {2010})}\BibitemShut {NoStop}%
\bibitem [{\citenamefont {Aikawa}\ \emph {et~al.}(2010)\citenamefont {Aikawa},
  \citenamefont {Akamatsu}, \citenamefont {Hayashi}, \citenamefont {Oasa},
  \citenamefont {Kobayashi}, \citenamefont {Naidon}, \citenamefont {Kishimoto},
  \citenamefont {Ueda},\ and\ \citenamefont {Inouye}}]{Aikawa:2010}%
  \BibitemOpen
  \bibfield  {author} {\bibinfo {author} {\bibfnamefont {K.}~\bibnamefont
  {Aikawa}}, \bibinfo {author} {\bibfnamefont {D.}~\bibnamefont {Akamatsu}},
  \bibinfo {author} {\bibfnamefont {M.}~\bibnamefont {Hayashi}}, \bibinfo
  {author} {\bibfnamefont {K.}~\bibnamefont {Oasa}}, \bibinfo {author}
  {\bibfnamefont {J.}~\bibnamefont {Kobayashi}}, \bibinfo {author}
  {\bibfnamefont {P.}~\bibnamefont {Naidon}}, \bibinfo {author} {\bibfnamefont
  {T.}~\bibnamefont {Kishimoto}}, \bibinfo {author} {\bibfnamefont
  {M.}~\bibnamefont {Ueda}}, \ and\ \bibinfo {author} {\bibfnamefont
  {S.}~\bibnamefont {Inouye}},\ }\href@noop {} {\bibfield  {journal} {\bibinfo
  {journal} {Phys. Rev. Lett.}\ }\textbf {\bibinfo {volume} {105}},\ \bibinfo
  {pages} {203001} (\bibinfo {year} {2010})}\BibitemShut {NoStop}%
\bibitem [{\citenamefont {Ospelkaus}\ \emph {et~al.}(2010)\citenamefont
  {Ospelkaus}, \citenamefont {Ni}, \citenamefont {Wang}, \citenamefont {{de
  Miranda}}, \citenamefont {Neyenhuis}, \citenamefont {Qu\'{e}m\'{e}ner},
  \citenamefont {Julienne}, \citenamefont {Bohn}, \citenamefont {Jin},\ and\
  \citenamefont {Ye}}]{Ospelkaus:react:2010}%
  \BibitemOpen
  \bibfield  {author} {\bibinfo {author} {\bibfnamefont {S.}~\bibnamefont
  {Ospelkaus}}, \bibinfo {author} {\bibfnamefont {K.-K.}\ \bibnamefont {Ni}},
  \bibinfo {author} {\bibfnamefont {D.}~\bibnamefont {Wang}}, \bibinfo {author}
  {\bibfnamefont {M.~H.~G.}\ \bibnamefont {{de Miranda}}}, \bibinfo {author}
  {\bibfnamefont {B.}~\bibnamefont {Neyenhuis}}, \bibinfo {author}
  {\bibfnamefont {G.}~\bibnamefont {Qu\'{e}m\'{e}ner}}, \bibinfo {author}
  {\bibfnamefont {P.~S.}\ \bibnamefont {Julienne}}, \bibinfo {author}
  {\bibfnamefont {J.~L.}\ \bibnamefont {Bohn}}, \bibinfo {author}
  {\bibfnamefont {D.~S.}\ \bibnamefont {Jin}}, \ and\ \bibinfo {author}
  {\bibfnamefont {J.}~\bibnamefont {Ye}},\ }\href@noop {} {\bibfield  {journal}
  {\bibinfo  {journal} {Science}\ }\textbf {\bibinfo {volume} {327}},\ \bibinfo
  {pages} {853} (\bibinfo {year} {2010})}\BibitemShut {NoStop}%
\bibitem [{\citenamefont {Ni}\ \emph {et~al.}(2010)\citenamefont {Ni},
  \citenamefont {Ospelkaus}, \citenamefont {Wang}, \citenamefont
  {Qu\'em\'ener}, \citenamefont {Neyenhuis}, \citenamefont {{de Miranda}},
  \citenamefont {Bohn}, \citenamefont {Ye},\ and\ \citenamefont
  {Jin}}]{Ni:2010}%
  \BibitemOpen
  \bibfield  {author} {\bibinfo {author} {\bibfnamefont {K.-K.}\ \bibnamefont
  {Ni}}, \bibinfo {author} {\bibfnamefont {S.}~\bibnamefont {Ospelkaus}},
  \bibinfo {author} {\bibfnamefont {D.}~\bibnamefont {Wang}}, \bibinfo {author}
  {\bibfnamefont {G.}~\bibnamefont {Qu\'em\'ener}}, \bibinfo {author}
  {\bibfnamefont {B.}~\bibnamefont {Neyenhuis}}, \bibinfo {author}
  {\bibfnamefont {M.~H.~G.}\ \bibnamefont {{de Miranda}}}, \bibinfo {author}
  {\bibfnamefont {J.~L.}\ \bibnamefont {Bohn}}, \bibinfo {author}
  {\bibfnamefont {J.}~\bibnamefont {Ye}}, \ and\ \bibinfo {author}
  {\bibfnamefont {D.~S.}\ \bibnamefont {Jin}},\ }\href@noop {} {\bibfield
  {journal} {\bibinfo  {journal} {Nature}\ }\textbf {\bibinfo {volume} {464}},\
  \bibinfo {pages} {1324} (\bibinfo {year} {2010})}\BibitemShut {NoStop}%
\bibitem [{\citenamefont {Micheli}\ \emph {et~al.}(2006)\citenamefont
  {Micheli}, \citenamefont {Brennen},\ and\ \citenamefont
  {Zoller}}]{Micheli:2006}%
  \BibitemOpen
  \bibfield  {author} {\bibinfo {author} {\bibfnamefont {A.}~\bibnamefont
  {Micheli}}, \bibinfo {author} {\bibfnamefont {G.~K.}\ \bibnamefont
  {Brennen}}, \ and\ \bibinfo {author} {\bibfnamefont {P.}~\bibnamefont
  {Zoller}},\ }\href@noop {} {\bibfield  {journal} {\bibinfo  {journal} {Nature
  Phys.}\ }\textbf {\bibinfo {volume} {2}},\ \bibinfo {pages} {341} (\bibinfo
  {year} {2006})}\BibitemShut {NoStop}%
\bibitem [{\citenamefont {\.Zuchowski}\ \emph {et~al.}(2010)\citenamefont
  {\.Zuchowski}, \citenamefont {Aldegunde},\ and\ \citenamefont
  {Hutson}}]{Zuchowski:RbSr:2010}%
  \BibitemOpen
  \bibfield  {author} {\bibinfo {author} {\bibfnamefont {P.~S.}\ \bibnamefont
  {\.Zuchowski}}, \bibinfo {author} {\bibfnamefont {J.}~\bibnamefont
  {Aldegunde}}, \ and\ \bibinfo {author} {\bibfnamefont {J.~M.}\ \bibnamefont
  {Hutson}},\ }\href@noop {} {\bibfield  {journal} {\bibinfo  {journal} {Phys.
  Rev. Lett.}\ }\textbf {\bibinfo {volume} {105}},\ \bibinfo {pages} {153201}
  (\bibinfo {year} {2010})}\BibitemShut {NoStop}%
\bibitem [{\citenamefont {Brue}\ and\ \citenamefont
  {Hutson}(2012)}]{Brue:LiYb:2012}%
  \BibitemOpen
  \bibfield  {author} {\bibinfo {author} {\bibfnamefont {D.~A.}\ \bibnamefont
  {Brue}}\ and\ \bibinfo {author} {\bibfnamefont {J.~M.}\ \bibnamefont
  {Hutson}},\ }\href@noop {} {\bibfield  {journal} {\bibinfo  {journal} {Phys.
  Rev. Lett.}\ }\textbf {\bibinfo {volume} {108}},\ \bibinfo {pages} {043201}
  (\bibinfo {year} {2012})}\BibitemShut {NoStop}%
\bibitem [{\citenamefont {Brue}\ and\ \citenamefont
  {Hutson}(2013)}]{Brue:AlkYb:2013}%
  \BibitemOpen
  \bibfield  {author} {\bibinfo {author} {\bibfnamefont {D.~A.}\ \bibnamefont
  {Brue}}\ and\ \bibinfo {author} {\bibfnamefont {J.~M.}\ \bibnamefont
  {Hutson}},\ }\href@noop {} {\bibfield  {journal} {\bibinfo  {journal} {Phys.
  Rev. A}\ }\textbf {\bibinfo {volume} {87}},\ \bibinfo {pages} {052709}
  (\bibinfo {year} {2013})}\BibitemShut {NoStop}%
\bibitem [{\citenamefont {Okano}\ \emph {et~al.}(2009)\citenamefont {Okano},
  \citenamefont {Hara}, \citenamefont {Muramatsu}, \citenamefont {Doi},
  \citenamefont {Uetake}, \citenamefont {Takasu},\ and\ \citenamefont
  {Takahashi}}]{Okano:2009}%
  \BibitemOpen
  \bibfield  {author} {\bibinfo {author} {\bibfnamefont {M.}~\bibnamefont
  {Okano}}, \bibinfo {author} {\bibfnamefont {H.}~\bibnamefont {Hara}},
  \bibinfo {author} {\bibfnamefont {M.}~\bibnamefont {Muramatsu}}, \bibinfo
  {author} {\bibfnamefont {K.}~\bibnamefont {Doi}}, \bibinfo {author}
  {\bibfnamefont {S.}~\bibnamefont {Uetake}}, \bibinfo {author} {\bibfnamefont
  {Y.}~\bibnamefont {Takasu}}, \ and\ \bibinfo {author} {\bibfnamefont
  {Y.}~\bibnamefont {Takahashi}},\ }\href@noop {} {\bibfield  {journal}
  {\bibinfo  {journal} {Appl. Phys. B}\ }\textbf {\bibinfo {volume} {98}},\
  \bibinfo {pages} {691} (\bibinfo {year} {2009})}\BibitemShut {NoStop}%
\bibitem [{\citenamefont {Hansen}\ \emph {et~al.}(2011)\citenamefont {Hansen},
  \citenamefont {Khramov}, \citenamefont {Dowd}, \citenamefont {Jamison},
  \citenamefont {Ivanov},\ and\ \citenamefont {Gupta}}]{Hansen:2011}%
  \BibitemOpen
  \bibfield  {author} {\bibinfo {author} {\bibfnamefont {A.~H.}\ \bibnamefont
  {Hansen}}, \bibinfo {author} {\bibfnamefont {A.~Y.}\ \bibnamefont {Khramov}},
  \bibinfo {author} {\bibfnamefont {W.~H.}\ \bibnamefont {Dowd}}, \bibinfo
  {author} {\bibfnamefont {A.~O.}\ \bibnamefont {Jamison}}, \bibinfo {author}
  {\bibfnamefont {V.~V.}\ \bibnamefont {Ivanov}}, \ and\ \bibinfo {author}
  {\bibfnamefont {S.}~\bibnamefont {Gupta}},\ }\href {\doibase
  10.1103/PhysRevA.84.011606} {\bibfield  {journal} {\bibinfo  {journal} {Phys.
  Rev. A}\ }\textbf {\bibinfo {volume} {84}},\ \bibinfo {pages} {011606}
  (\bibinfo {year} {2011})}\BibitemShut {NoStop}%
\bibitem [{\citenamefont {Ivanov}\ \emph {et~al.}(2011)\citenamefont {Ivanov},
  \citenamefont {Khramov}, \citenamefont {Hansen}, \citenamefont {Dowd},
  \citenamefont {M\"unchow}, \citenamefont {Jamison},\ and\ \citenamefont
  {Gupta}}]{Ivanov:2011}%
  \BibitemOpen
  \bibfield  {author} {\bibinfo {author} {\bibfnamefont {V.~V.}\ \bibnamefont
  {Ivanov}}, \bibinfo {author} {\bibfnamefont {A.~Y.}\ \bibnamefont {Khramov}},
  \bibinfo {author} {\bibfnamefont {A.~H.}\ \bibnamefont {Hansen}}, \bibinfo
  {author} {\bibfnamefont {W.~H.}\ \bibnamefont {Dowd}}, \bibinfo {author}
  {\bibfnamefont {F.}~\bibnamefont {M\"unchow}}, \bibinfo {author}
  {\bibfnamefont {A.~O.}\ \bibnamefont {Jamison}}, \ and\ \bibinfo {author}
  {\bibfnamefont {S.}~\bibnamefont {Gupta}},\ }\href@noop {} {\bibfield
  {journal} {\bibinfo  {journal} {Phys. Rev. Lett.}\ }\textbf {\bibinfo
  {volume} {106}},\ \bibinfo {pages} {153201} (\bibinfo {year}
  {2011})}\BibitemShut {NoStop}%
\bibitem [{\citenamefont {Baumer}\ \emph {et~al.}(2011)\citenamefont {Baumer},
  \citenamefont {M\"unchow}, \citenamefont {G\"orlitz}, \citenamefont
  {Maxwell}, \citenamefont {Julienne},\ and\ \citenamefont
  {Tiesinga}}]{Baumer:2011}%
  \BibitemOpen
  \bibfield  {author} {\bibinfo {author} {\bibfnamefont {F.}~\bibnamefont
  {Baumer}}, \bibinfo {author} {\bibfnamefont {F.}~\bibnamefont {M\"unchow}},
  \bibinfo {author} {\bibfnamefont {A.}~\bibnamefont {G\"orlitz}}, \bibinfo
  {author} {\bibfnamefont {S.~E.}\ \bibnamefont {Maxwell}}, \bibinfo {author}
  {\bibfnamefont {P.~S.}\ \bibnamefont {Julienne}}, \ and\ \bibinfo {author}
  {\bibfnamefont {E.}~\bibnamefont {Tiesinga}},\ }\href {\doibase
  10.1103/PhysRevA.83.040702} {\bibfield  {journal} {\bibinfo  {journal} {Phys.
  Rev. A}\ }\textbf {\bibinfo {volume} {83}},\ \bibinfo {pages} {040702}
  (\bibinfo {year} {2011})}\BibitemShut {NoStop}%
\bibitem [{\citenamefont {M\"unchow}\ \emph {et~al.}(2011)\citenamefont
  {M\"unchow}, \citenamefont {Bruni}, \citenamefont {Madalinskia},\ and\
  \citenamefont {G\"orlitz}}]{Muenchow:2011}%
  \BibitemOpen
  \bibfield  {author} {\bibinfo {author} {\bibfnamefont {F.}~\bibnamefont
  {M\"unchow}}, \bibinfo {author} {\bibfnamefont {C.}~\bibnamefont {Bruni}},
  \bibinfo {author} {\bibfnamefont {M.}~\bibnamefont {Madalinskia}}, \ and\
  \bibinfo {author} {\bibfnamefont {A.}~\bibnamefont {G\"orlitz}},\ }\href@noop
  {} {\bibfield  {journal} {\bibinfo  {journal} {Phys. Chem. Chem. Phys.}\
  }\textbf {\bibinfo {volume} {13}},\ \bibinfo {pages} {18734} (\bibinfo {year}
  {2011})}\BibitemShut {NoStop}%
\bibitem [{\citenamefont {M\"unchow}(2012)}]{Muenchow:thesis:2012}%
  \BibitemOpen
  \bibfield  {author} {\bibinfo {author} {\bibfnamefont {F.}~\bibnamefont
  {M\"unchow}},\ }\emph {\bibinfo {title} {2-photon photoassociation
  spectroscopy in a mixture of {Y}tterbium and {R}ubidium}},\ \href@noop {}
  {Ph.D. thesis},\ \bibinfo  {school} {Heinrich-Heine-Universit\"at}, \bibinfo
  {address} {D\"usseldorf} (\bibinfo {year} {2012})\BibitemShut {NoStop}%
\bibitem [{\citenamefont {Yamaguchi}\ \emph {et~al.}(2008)\citenamefont
  {Yamaguchi}, \citenamefont {Uetake}, \citenamefont {Hashimoto}, \citenamefont
  {Doyle},\ and\ \citenamefont {Takahashi}}]{Yamaguchi:2008}%
  \BibitemOpen
  \bibfield  {author} {\bibinfo {author} {\bibfnamefont {A.}~\bibnamefont
  {Yamaguchi}}, \bibinfo {author} {\bibfnamefont {S.}~\bibnamefont {Uetake}},
  \bibinfo {author} {\bibfnamefont {D.}~\bibnamefont {Hashimoto}}, \bibinfo
  {author} {\bibfnamefont {J.~M.}\ \bibnamefont {Doyle}}, \ and\ \bibinfo
  {author} {\bibfnamefont {Y.}~\bibnamefont {Takahashi}},\ }\href@noop {}
  {\bibfield  {journal} {\bibinfo  {journal} {Phys. Rev. Lett.}\ }\textbf
  {\bibinfo {volume} {101}},\ \bibinfo {pages} {233002} (\bibinfo {year}
  {2008})}\BibitemShut {NoStop}%
\bibitem [{\citenamefont {Mishra}\ and\ \citenamefont
  {Balasubramanian}(2001)}]{Mishra:2001}%
  \BibitemOpen
  \bibfield  {author} {\bibinfo {author} {\bibfnamefont {A.~P.}\ \bibnamefont
  {Mishra}}\ and\ \bibinfo {author} {\bibfnamefont {T.~K.}\ \bibnamefont
  {Balasubramanian}},\ }\href@noop {} {\bibfield  {journal} {\bibinfo
  {journal} {J. Quant. Spectrosc. Rad. Transf.}\ }\textbf {\bibinfo {volume}
  {69}},\ \bibinfo {pages} {760} (\bibinfo {year} {2001})}\BibitemShut
  {NoStop}%
\bibitem [{\citenamefont {Reid}\ and\ \citenamefont
  {Dalgarno}(1969)}]{Reid:1969}%
  \BibitemOpen
  \bibfield  {author} {\bibinfo {author} {\bibfnamefont {R.~H.~G.}\
  \bibnamefont {Reid}}\ and\ \bibinfo {author} {\bibfnamefont {A.}~\bibnamefont
  {Dalgarno}},\ }\href@noop {} {\bibfield  {journal} {\bibinfo  {journal}
  {Phys. Rev. Lett.}\ }\textbf {\bibinfo {volume} {22}},\ \bibinfo {pages}
  {1029} (\bibinfo {year} {1969})}\BibitemShut {NoStop}%
\bibitem [{\citenamefont {Krems}\ \emph {et~al.}(2004)\citenamefont {Krems},
  \citenamefont {Groenenboom},\ and\ \citenamefont
  {Dalgarno}}]{Krems:atoms:2004}%
  \BibitemOpen
  \bibfield  {author} {\bibinfo {author} {\bibfnamefont {R.~V.}\ \bibnamefont
  {Krems}}, \bibinfo {author} {\bibfnamefont {G.~C.}\ \bibnamefont
  {Groenenboom}}, \ and\ \bibinfo {author} {\bibfnamefont {A.}~\bibnamefont
  {Dalgarno}},\ }\href@noop {} {\bibfield  {journal} {\bibinfo  {journal} {J.
  Phys. Chem. A}\ }\textbf {\bibinfo {volume} {108}},\ \bibinfo {pages} {8941}
  (\bibinfo {year} {2004})}\BibitemShut {NoStop}%
\bibitem [{\citenamefont {Hansen}\ \emph {et~al.}(2013)\citenamefont {Hansen},
  \citenamefont {Khramov}, \citenamefont {Dowd}, \citenamefont {Jamison},
  \citenamefont {Plotkin-Swing}, \citenamefont {Roy},\ and\ \citenamefont
  {Gupta}}]{Hansen:2013}%
  \BibitemOpen
  \bibfield  {author} {\bibinfo {author} {\bibfnamefont {A.~H.}\ \bibnamefont
  {Hansen}}, \bibinfo {author} {\bibfnamefont {A.~Y.}\ \bibnamefont {Khramov}},
  \bibinfo {author} {\bibfnamefont {W.~H.}\ \bibnamefont {Dowd}}, \bibinfo
  {author} {\bibfnamefont {A.~O.}\ \bibnamefont {Jamison}}, \bibinfo {author}
  {\bibfnamefont {B.}~\bibnamefont {Plotkin-Swing}}, \bibinfo {author}
  {\bibfnamefont {R.~J.}\ \bibnamefont {Roy}}, \ and\ \bibinfo {author}
  {\bibfnamefont {S.}~\bibnamefont {Gupta}},\ }\href@noop {} {\bibfield
  {journal} {\bibinfo  {journal} {Phys. Rev. A}\ }\textbf {\bibinfo {volume}
  {87}},\ \bibinfo {pages} {013615} (\bibinfo {year} {2013})}\BibitemShut
  {NoStop}%
\bibitem [{\citenamefont {Gopakumar}\ \emph {et~al.}(2010)\citenamefont
  {Gopakumar}, \citenamefont {Abe}, \citenamefont {Das}, \citenamefont {Hada},\
  and\ \citenamefont {Hirao}}]{Gopakumar:2010}%
  \BibitemOpen
  \bibfield  {author} {\bibinfo {author} {\bibfnamefont {G.}~\bibnamefont
  {Gopakumar}}, \bibinfo {author} {\bibfnamefont {M.}~\bibnamefont {Abe}},
  \bibinfo {author} {\bibfnamefont {B.~P.}\ \bibnamefont {Das}}, \bibinfo
  {author} {\bibfnamefont {M.}~\bibnamefont {Hada}}, \ and\ \bibinfo {author}
  {\bibfnamefont {K.}~\bibnamefont {Hirao}},\ }\href@noop {} {\bibfield
  {journal} {\bibinfo  {journal} {J. Chem. Phys.}\ }\textbf {\bibinfo {volume}
  {133}},\ \bibinfo {pages} {124317} (\bibinfo {year} {2010})}\BibitemShut
  {NoStop}%
\bibitem [{\citenamefont {Zhang}\ \emph {et~al.}(2010)\citenamefont {Zhang},
  \citenamefont {Sadeghpour},\ and\ \citenamefont {Dalgarno}}]{Zhang:2010}%
  \BibitemOpen
  \bibfield  {author} {\bibinfo {author} {\bibfnamefont {P.}~\bibnamefont
  {Zhang}}, \bibinfo {author} {\bibfnamefont {H.~R.}\ \bibnamefont
  {Sadeghpour}}, \ and\ \bibinfo {author} {\bibfnamefont {A.}~\bibnamefont
  {Dalgarno}},\ }\href@noop {} {\bibfield  {journal} {\bibinfo  {journal} {J.
  Chem. Phys.}\ }\textbf {\bibinfo {volume} {133}},\ \bibinfo {pages} {044306}
  (\bibinfo {year} {2010})}\BibitemShut {NoStop}%
\bibitem [{\citenamefont {Ho}\ and\ \citenamefont {Rabitz}(1996)}]{Ho:1996}%
  \BibitemOpen
  \bibfield  {author} {\bibinfo {author} {\bibfnamefont {T.~S.}\ \bibnamefont
  {Ho}}\ and\ \bibinfo {author} {\bibfnamefont {H.}~\bibnamefont {Rabitz}},\
  }\href@noop {} {\bibfield  {journal} {\bibinfo  {journal} {J. Chem. Phys.}\
  }\textbf {\bibinfo {volume} {104}},\ \bibinfo {pages} {2584} (\bibinfo {year}
  {1996})}\BibitemShut {NoStop}%
\bibitem [{\citenamefont {Tang}(1969)}]{Tang:1969}%
  \BibitemOpen
  \bibfield  {author} {\bibinfo {author} {\bibfnamefont {K.~T.}\ \bibnamefont
  {Tang}},\ }\href {\doibase 10.1103/PhysRev.177.108} {\bibfield  {journal}
  {\bibinfo  {journal} {Phys. Rev.}\ }\textbf {\bibinfo {volume} {177}},\
  \bibinfo {pages} {108} (\bibinfo {year} {1969})}\BibitemShut {NoStop}%
\bibitem [{\citenamefont {Derevianko}\ \emph {et~al.}(2010)\citenamefont
  {Derevianko}, \citenamefont {Porsev},\ and\ \citenamefont
  {Babb}}]{Derevianko:2010}%
  \BibitemOpen
  \bibfield  {author} {\bibinfo {author} {\bibfnamefont {A.}~\bibnamefont
  {Derevianko}}, \bibinfo {author} {\bibfnamefont {S.~G.}\ \bibnamefont
  {Porsev}}, \ and\ \bibinfo {author} {\bibfnamefont {J.~F.}\ \bibnamefont
  {Babb}},\ }\href@noop {} {\bibfield  {journal} {\bibinfo  {journal} {Atomic
  Data and Nuclear Data Tables}\ }\textbf {\bibinfo {volume} {96}},\ \bibinfo
  {pages} {323} (\bibinfo {year} {2010})}\BibitemShut {NoStop}%
\bibitem [{\citenamefont {Dzuba}\ and\ \citenamefont
  {Derevianko}(2010)}]{Dzuba:2010}%
  \BibitemOpen
  \bibfield  {author} {\bibinfo {author} {\bibfnamefont {V.~A.}\ \bibnamefont
  {Dzuba}}\ and\ \bibinfo {author} {\bibfnamefont {A.}~\bibnamefont
  {Derevianko}},\ }\href@noop {} {\bibfield  {journal} {\bibinfo  {journal} {J.
  Phys. B}\ }\textbf {\bibinfo {volume} {43}},\ \bibinfo {pages} {074011}
  (\bibinfo {year} {2010})}\BibitemShut {NoStop}%
\bibitem [{\citenamefont {Jiang}\ \emph {et~al.}(2013)\citenamefont {Jiang},
  \citenamefont {Cheng},\ and\ \citenamefont {Mitroy}}]{Jiang:2013}%
  \BibitemOpen
  \bibfield  {author} {\bibinfo {author} {\bibfnamefont {J.}~\bibnamefont
  {Jiang}}, \bibinfo {author} {\bibfnamefont {Y.}~\bibnamefont {Cheng}}, \ and\
  \bibinfo {author} {\bibfnamefont {J.}~\bibnamefont {Mitroy}},\ }\href@noop {}
  {\bibfield  {journal} {\bibinfo  {journal} {arXiv:1303.1234}\ } (\bibinfo
  {year} {2013})}\BibitemShut {NoStop}%
\bibitem [{Note1()}]{Note1}%
  \BibitemOpen
  \bibinfo {note} {To obtain smooth curves with the correct long-range
  behavior, we omitted {\protect \em ab initio} points outside $R=13.0, 8.6,
  9.6$ and $8.6\protect \,a_0$ for the $^2\Sigma ^+$, $^2\Pi $, $^4\Sigma ^+$
  and $^4\Pi $ curves, respectively, and between 9.6 and $10.5\protect \,a_0$
  for the $^2\Sigma ^+$ curve.}\BibitemShut {Stop}%
\bibitem [{\citenamefont {Meggers}\ and\ \citenamefont
  {Tech}(1978)}]{Meggers:1978}%
  \BibitemOpen
  \bibfield  {author} {\bibinfo {author} {\bibfnamefont {W.~F.}\ \bibnamefont
  {Meggers}}\ and\ \bibinfo {author} {\bibfnamefont {J.~L.}\ \bibnamefont
  {Tech}},\ }\href@noop {} {\bibfield  {journal} {\bibinfo  {journal} {J. Res.
  Natl. Bur. Stand. (U.S.)}\ }\textbf {\bibinfo {volume} {83}},\ \bibinfo
  {pages} {13} (\bibinfo {year} {1978})}\BibitemShut {NoStop}%
\bibitem [{\citenamefont {Hutson}(2011)}]{Hutson:bound:2011}%
  \BibitemOpen
  \bibfield  {author} {\bibinfo {author} {\bibfnamefont {J.~M.}\ \bibnamefont
  {Hutson}},\ }\href@noop {} {} (\bibinfo {year} {2011}),\ \bibinfo {note}
  {{BOUND} computer code}\BibitemShut {NoStop}%
\bibitem [{\citenamefont {Hutson}\ and\ \citenamefont
  {Green}(1994)}]{molscat:v14-short}%
  \BibitemOpen
  \bibfield  {author} {\bibinfo {author} {\bibfnamefont {J.~M.}\ \bibnamefont
  {Hutson}}\ and\ \bibinfo {author} {\bibfnamefont {S.}~\bibnamefont {Green}},\
  }\href@noop {} {\emph {\bibinfo {title} {{MOLSCAT} computer program, version
  14}}}\ (\bibinfo  {publisher} {CCP6},\ \bibinfo {address} {Daresbury},\
  \bibinfo {year} {1994})\BibitemShut {NoStop}%
\bibitem [{\citenamefont {Hutson}(2007)}]{Hutson:res:2007}%
  \BibitemOpen
  \bibfield  {author} {\bibinfo {author} {\bibfnamefont {J.~M.}\ \bibnamefont
  {Hutson}},\ }\href@noop {} {\bibfield  {journal} {\bibinfo  {journal} {New J.
  Phys.}\ }\textbf {\bibinfo {volume} {9}},\ \bibinfo {pages} {152} (\bibinfo
  {year} {2007})}\BibitemShut {NoStop}%
\bibitem [{\citenamefont {Idziaszek}\ and\ \citenamefont
  {Julienne}(2010)}]{Idziaszek:PRL:2010}%
  \BibitemOpen
  \bibfield  {author} {\bibinfo {author} {\bibfnamefont {Z.}~\bibnamefont
  {Idziaszek}}\ and\ \bibinfo {author} {\bibfnamefont {P.~S.}\ \bibnamefont
  {Julienne}},\ }\href@noop {} {\bibfield  {journal} {\bibinfo  {journal}
  {Phys. Rev. Lett.}\ }\textbf {\bibinfo {volume} {104}},\ \bibinfo {pages}
  {113202} (\bibinfo {year} {2010})}\BibitemShut {NoStop}%
\bibitem [{\citenamefont {Uetake}\ \emph {et~al.}(2012)\citenamefont {Uetake},
  \citenamefont {Murakami}, \citenamefont {Doyle},\ and\ \citenamefont
  {Takahashi}}]{Uetake:2012}%
  \BibitemOpen
  \bibfield  {author} {\bibinfo {author} {\bibfnamefont {S.}~\bibnamefont
  {Uetake}}, \bibinfo {author} {\bibfnamefont {R.}~\bibnamefont {Murakami}},
  \bibinfo {author} {\bibfnamefont {J.~M.}\ \bibnamefont {Doyle}}, \ and\
  \bibinfo {author} {\bibfnamefont {Y.}~\bibnamefont {Takahashi}},\ }\href@noop
  {} {\bibfield  {journal} {\bibinfo  {journal} {Phys. Rev. A}\ }\textbf
  {\bibinfo {volume} {86}},\ \bibinfo {pages} {032712} (\bibinfo {year}
  {2012})}\BibitemShut {NoStop}%
\bibitem [{\citenamefont {Julienne}\ \emph {et~al.}(1997)\citenamefont
  {Julienne}, \citenamefont {Mies}, \citenamefont {Tiesinga},\ and\
  \citenamefont {Williams}}]{Julienne:1997}%
  \BibitemOpen
  \bibfield  {author} {\bibinfo {author} {\bibfnamefont {P.~S.}\ \bibnamefont
  {Julienne}}, \bibinfo {author} {\bibfnamefont {F.~H.}\ \bibnamefont {Mies}},
  \bibinfo {author} {\bibfnamefont {E.}~\bibnamefont {Tiesinga}}, \ and\
  \bibinfo {author} {\bibfnamefont {C.~J.}\ \bibnamefont {Williams}},\
  }\href@noop {} {\bibfield  {journal} {\bibinfo  {journal} {Phys. Rev. Lett.}\
  }\textbf {\bibinfo {volume} {78}},\ \bibinfo {pages} {1880} (\bibinfo {year}
  {1997})}\BibitemShut {NoStop}%
\bibitem [{\citenamefont {Gribakin}\ and\ \citenamefont
  {Flambaum}(1993)}]{Gribakin:1993}%
  \BibitemOpen
  \bibfield  {author} {\bibinfo {author} {\bibfnamefont {G.~F.}\ \bibnamefont
  {Gribakin}}\ and\ \bibinfo {author} {\bibfnamefont {V.~V.}\ \bibnamefont
  {Flambaum}},\ }\href@noop {} {\bibfield  {journal} {\bibinfo  {journal}
  {Phys. Rev. A}\ }\textbf {\bibinfo {volume} {48}},\ \bibinfo {pages} {546}
  (\bibinfo {year} {1993})}\BibitemShut {NoStop}%
\bibitem [{\citenamefont {Porsev}\ \emph {et~al.}(2004)\citenamefont {Porsev},
  \citenamefont {Derevianko},\ and\ \citenamefont {Fortson}}]{Porsev:2004}%
  \BibitemOpen
  \bibfield  {author} {\bibinfo {author} {\bibfnamefont {S.~G.}\ \bibnamefont
  {Porsev}}, \bibinfo {author} {\bibfnamefont {A.}~\bibnamefont {Derevianko}},
  \ and\ \bibinfo {author} {\bibfnamefont {E.~N.}\ \bibnamefont {Fortson}},\
  }\href {\doibase 10.1103/PhysRevA.69.021403} {\bibfield  {journal} {\bibinfo
  {journal} {Phys. Rev. A}\ }\textbf {\bibinfo {volume} {69}},\ \bibinfo
  {pages} {021403} (\bibinfo {year} {2004})}\BibitemShut {NoStop}%
\bibitem [{\citenamefont {Hoyt}\ \emph {et~al.}(2005)\citenamefont {Hoyt},
  \citenamefont {Barber}, \citenamefont {Oates}, \citenamefont {Fortier},
  \citenamefont {Diddams},\ and\ \citenamefont {Hollberg}}]{Hoyt:2005}%
  \BibitemOpen
  \bibfield  {author} {\bibinfo {author} {\bibfnamefont {C.~W.}\ \bibnamefont
  {Hoyt}}, \bibinfo {author} {\bibfnamefont {Z.~W.}\ \bibnamefont {Barber}},
  \bibinfo {author} {\bibfnamefont {C.~W.}\ \bibnamefont {Oates}}, \bibinfo
  {author} {\bibfnamefont {T.~M.}\ \bibnamefont {Fortier}}, \bibinfo {author}
  {\bibfnamefont {S.~A.}\ \bibnamefont {Diddams}}, \ and\ \bibinfo {author}
  {\bibfnamefont {L.}~\bibnamefont {Hollberg}},\ }\href {\doibase
  10.1103/PhysRevLett.95.083003} {\bibfield  {journal} {\bibinfo  {journal}
  {Phys. Rev. Lett.}\ }\textbf {\bibinfo {volume} {95}},\ \bibinfo {pages}
  {083003} (\bibinfo {year} {2005})}\BibitemShut {NoStop}%
\bibitem [{\citenamefont {Hong}\ \emph {et~al.}(2005)\citenamefont {Hong},
  \citenamefont {Cramer}, \citenamefont {Cook}, \citenamefont {Nagourney},\
  and\ \citenamefont {Fortson}}]{Hong:2005}%
  \BibitemOpen
  \bibfield  {author} {\bibinfo {author} {\bibfnamefont {T.}~\bibnamefont
  {Hong}}, \bibinfo {author} {\bibfnamefont {C.}~\bibnamefont {Cramer}},
  \bibinfo {author} {\bibfnamefont {E.}~\bibnamefont {Cook}}, \bibinfo {author}
  {\bibfnamefont {W.}~\bibnamefont {Nagourney}}, \ and\ \bibinfo {author}
  {\bibfnamefont {E.~N.}\ \bibnamefont {Fortson}},\ }\href@noop {} {\bibfield
  {journal} {\bibinfo  {journal} {Opt. Lett.}\ }\textbf {\bibinfo {volume}
  {30}},\ \bibinfo {pages} {2644} (\bibinfo {year} {2005})}\BibitemShut
  {NoStop}%
\bibitem [{\citenamefont {Barber}\ \emph {et~al.}(2006)\citenamefont {Barber},
  \citenamefont {Hoyt}, \citenamefont {Oates}, \citenamefont {Hollberg},
  \citenamefont {Taichenachev},\ and\ \citenamefont {Yudin}}]{Barber:2006}%
  \BibitemOpen
  \bibfield  {author} {\bibinfo {author} {\bibfnamefont {Z.~W.}\ \bibnamefont
  {Barber}}, \bibinfo {author} {\bibfnamefont {C.~W.}\ \bibnamefont {Hoyt}},
  \bibinfo {author} {\bibfnamefont {C.~W.}\ \bibnamefont {Oates}}, \bibinfo
  {author} {\bibfnamefont {L.}~\bibnamefont {Hollberg}}, \bibinfo {author}
  {\bibfnamefont {A.~V.}\ \bibnamefont {Taichenachev}}, \ and\ \bibinfo
  {author} {\bibfnamefont {V.~I.}\ \bibnamefont {Yudin}},\ }\href {\doibase
  10.1103/PhysRevLett.96.083002} {\bibfield  {journal} {\bibinfo  {journal}
  {Phys. Rev. Lett.}\ }\textbf {\bibinfo {volume} {96}},\ \bibinfo {pages}
  {083002} (\bibinfo {year} {2006})}\BibitemShut {NoStop}%
\bibitem [{\citenamefont {Kotochigova}(2013)}]{Kotochigova:private:2013}%
  \BibitemOpen
  \bibfield  {author} {\bibinfo {author} {\bibfnamefont {S.}~\bibnamefont
  {Kotochigova}},\ }\href@noop {} {} (\bibinfo {year} {2013}),\ \bibinfo {note}
  {private communication}\BibitemShut {NoStop}%
\end{thebibliography}%
\end{document}